\documentclass{elsarticle}
\usepackage{amssymb}
\usepackage{epsfig,graphicx}
\usepackage{graphicx}
\usepackage{amsmath}
\usepackage{color}

\newcommand{\be}{\begin{equation}}
\newcommand{\ee}{\end{equation}}

\newcommand{\bee}{\begin{eqnarray}}
\newcommand{\eee}{\end{eqnarray}}

\def\sp {\!+\!}
\def\sm {\!-\!}

\def\be{\begin{eqnarray} &&}

\def\ee{\end{eqnarray}}
\def\bew{\begin{widetext}}
\def\ew{\end{widetext}}

\usepackage{epsfig}
\def\lp {\left( }
\def\rp {\right) }
\def\lb {\left[ }
\def\rb {\right] }

\def\ra {\rangle }
\def\la {\langle }

\def\bea{\begin{eqnarray}}
\def\eea{\end{eqnarray}}
\def\nn {\nonumber}
\def\ni {\noindent}

\def\e{\epsilon}

\def\D {\Delta}

\def\m{\mu}

\def\p {\pi}

\newcommand{\bppp}{B^\pm \to  \pi^-\pi^+\pi^\pm}
\newcommand{\bkkk}{B^\pm \to  K^-K^+K^\pm}
\newcommand{\bkkp}{B^\pm \to  K^-K^+\pi^\pm}
\newcommand{\bkpp}{B^\pm \to  K^\pm\pi^-\pi^+}
\newcommand{\bddp}{B_c^\pm \to  D\bar D\pi^\pm}
\newcommand{\bddk}{B^\pm \to  D\bar DK^\pm}

\newcommand{\ddkk}{D\bar{D} \to K^+K^-}
\newcommand{\ddpp}{D\bar{D}\to \p\p}
\newcommand{\bckkp}{B_c^+ \to  K^- K^+ \pi^+}

\begin{document}
 \makeatletter
\begin{frontmatter}
\title{CP asymmetry from hadronic charm rescattering in  $\bppp$ decays  at the high mass region }
\author[CBPF]{I. Bediaga} 
\author[ITA]{T. Frederico}
\author[UB]{P. C. Magalh\~aes}
\ead{p.magalhaes@bristol.ac.uk}
\address[CBPF]{Centro Brasileiro de Pesquisas F\'isicas, 
22.290-180, Rio de Janeiro, RJ, Brazil}
\address[ITA]{Instituto Tecnol\'ogico de
Aeron\'autica, 12.228-900 S\~ao Jos\'e dos Campos, SP,
Brazil.}
\address[UB]{H.H. Wills Physics Laboratory, University of Bristol,  Bristol, BS8 1TLB, United Kingdom.  }
\date{\today}
\begin{abstract}

A model for the $\bppp$ decay amplitude is proposed to study the large CP violation observed  at the high mass region 
of the Dalitz plane. A short distance $ b \to u $ amplitude with the weak 
phase $\gamma$  is considered together with the contribution of a hadronic charm loop and a s-wave $\ddpp$ rescattering. In the model, the $\chi_{c0}$ appears as a narrow resonant 
state of the $D\bar D$ system below threshold. 
It is introduced in an unitary two channel S-matrix model of the coupled $D\bar D$ and $\pi\pi$ channels, where the $\chi_{c0}$ complex pole in 
$D\bar D$ channel shows its signature in the off-diagonal matrix element and in the associated $\ddpp$ transition amplitude.   
The strong phase of the resulting decay amplitude has a sharp sign change  at the  $D\bar D$ threshold, 
changing the sign of the CP asymmetry, as it is observed in the data. We conclude that  the hadronic charm loop and rescattering 
mechanism are  relevant to the broadening of the CP asymmetry around the $\chi_{c0}$  resonance in the $\pi\pi$ channel. This novel mechanism  provides a possible interpretation for the CP asymmetry challenging experimental result 
presented  by the LHCb collaboration for the $\bppp$ decay in the high mass region.
\end{abstract}
\end{frontmatter}

Experimental results from charmless three-body B decays  have shown a rich  distribution of CP violation (CPV) 
 within the Dalitz phase-space, the so called  Mirandizing distribution\footnote{CP asymmetry distribution in a Dalitz plot~\cite{BedPRD2009}.}~\cite{LHCb1,LHCb2,LHCb3}.  Positive and negative CP asymmetry are frequently seen in the same B decay channel and sometimes very close to each other in the phase-space, as have been observed in $\bkpp$ and $\bppp$ decays at low $\p\p$ mass. These particular phenomena can be explained through the interference term between the $\sigma$ and the $\rho(770)$ resonances \cite{LHCb3,ABAOT}. Another important source of CP asymmetry comes from the $\pi\pi\leftrightarrow KK$ rescattering, which couples different decay channels, namely, $\bkpp$ with $\bkkk$ and also   $\bkkp$  with  $\bppp$~\cite{LHCb1,LHCb2,LHCb3,ABAOT,BOT}.  
 Between others, these two sources of CP violation were already confirmed for $\bppp$~\cite{BpppPRD,BpppPRL} and $\bkkp$ decay channels~\cite{BkkpPRL} through a recent amplitude analysis performed by the LHCb collaboration for the Run I data.  There are also strong experimental evidences for CPV in the  Dalitz phase-space along the high mass region in all of those charged charmless B  three-body decays~\cite{LHCb3}. Although the source for this CPV is not yet identified, we can assume that the variation of the CP asymmetry  in the Dalitz plane is  originated by a running strong phase along the phase-space, since the dominant weak phase contributing to these decays is the CKM phase $\gamma$, which must be a constant.  

 Recently we proposed a new source of strong phase variation, associated with the possible $\ddkk$ rescattering, which couples the $\bddk$ to $\bkkk$  and the $\bddp$   to  $\bckkp$ decay channels~\cite{Pat2017, Pat2018}.  Where in the later we have also considered the contribution of the channel $B_c^\pm \to  D\bar D_s K^\pm$ through $D\bar{D_s}\to K\pi$ rescattering. In these studies, we concluded that  the long distance hadronic loop originated by the double charm penguin contribution can produce a strong  phase that changes along the Dalitz phase-space. 
The phase starts   around -1 radian  until the $D\bar{D}$ threshold, then it has a quick  phase variation given by  a sharp change from negative to positive values. 
 This phase variation can be responsible to change the sign of the CP asymmetry, as observed in experimental data~\cite{LHCb3}\footnote{\tiny{see  LHCb-PAPER-2014-044 supplemental material  at https://cds.cern.ch/record/1751517/files/.} }, if the associated amplitude is interfering with another one carrying  the weak phase.  
     
    More than two decades ago it was predicted  a CP violation
 in the high mass region of the $\bppp$ decay phase-space due to the presence of $\chi_{c0}$ resonance~\cite{Gronau96}.  $\chi_{c0}$ would be produced from $ b \to c \bar c d $ transition  at 
     tree level, without weak phase, and it can interfere with  $ b \to u \bar u d $  tree diagram amplitude, with weak phase,  leading to a strong CP asymmetry in the phase-space, with the possibility to extract the CKM weak phase $\gamma$~\cite{Gronau96,Gobel98}. One expects the $\chi_{c0}$  would be finally observed soon with the Run II LHCb data.  This conclusion is based on counting  the number of events already seen in the Cabibbo allowed  $B^\pm \to K^\pm \chi_{c0}$ decay, in the $\bkkk$ and  $\bkpp$ decays. Amplitude analysis performed by the Babar collaboration found a fit fraction of 1\% for these three-body final states \cite{BaBar_bkpipi} and \cite{BaBar_B3K}, respectively. From that  one can do a simple relation with  these decays    and the Cabibbo suppressed  $B^\pm \to \pi^\pm \chi_{c0}$ ($\frac{sin^2\theta}{cos^2\theta}\approx 0.05$) to estimate the number of events expected in LHCb Run II  for the $\bppp$ decay, arriving up to a few hundred events involving this scalar charmonium resonance.  
          
 Although LHCb did not find yet contribution from  the $B^\pm \to \pi^\pm \chi_{c0}$ amplitude in   $\bppp$ decay~\cite{BpppPRD,BpppPRL},  the Mirandizing 
distribution for Run I data~\cite{LHCb3} have shown already a clear and huge CP asymmetry around the  $\chi_{c0}$ invariant mass. This asymmetry suggests 
the presence of this resonance interfering with the nonresonant amplitude placed in this region\footnote{ Small amplitudes can be observed in the Dalitz plot when they interfere  with large ones, even before their peculiar signature becomes clear.}. However, 
the distribution of  CP asymmetry is much larger than the narrow width expected for this resonance, suggesting that part of the nonresonant background 
around the $\chi_{c0}$ peak comes from the same physical process that produces this resonance. 
Also, it is observed a change of sign in the CP asymmetry around the $D \bar D$ threshold, that can be assigned to the amplitude proposed in  \cite{Pat2017}.

The discussion about the importance of charm loops as a source of important 
contribution in heavy decay processes is not new~\cite{Wolfenstein,Soni2005,Colangelo1,Colangelo2,Hou2019}. 
In particular,  Colangelo et al.~\cite{Colangelo1,Colangelo2}  calculated the $B^-\to  K^- \chi_{c0}$ decay rate using a  hadronic triangle loop 
in combination with QCD factorization (QCDF) approach and HM$\chi$PT to describe the heavy-light mesons vertices,  including the coupling between 
$D\bar{D} \to c\bar{c}$ resonances. They argue that only QCDF cannot predict correctly the experimental branching fractions of the $B\to K (c\bar{c})$ transition, 
and in this framework the $B^-\to K^ - \chi_{c0}$ process is not allowed.
Indeed other models of the $\bppp$ decay amplitude were proposed in the literature using QCDF approach and none of them 
included $\chi_{c0}$~\cite{ManeG,KeriManeG,chinos,soni2007}. Contemporary to the present  work,  Ref. \cite{KV2020}
also  assumes that the  open-charm threshold may play an  important  role in generating CP violation. 
They used   the isobar approach in which  only  resonances  
above  threshold are considered and dressed by the coupling to the  $D\bar{D}$  intermediate state.

In this work, we explore the same mechanism used to describe the $\bkkk$ decays~\cite{Pat2017} 
 (also applied to the rare $B^+_c \to K^+ K^- \pi^+$ decay~\cite{Pat2018}), namely, the hadronic charm loop and $D\bar D$ 
 rescattering to  light pseudoscalars, to investigate the $\bppp$ decay, in an attempt to 
 extract the main qualitative features observed in the high mass region ($M^2_{\pi\pi} > 3$ GeV$^2$) of the CPV Mirandizing data distribution. The present study brings one important novelty to
the S-matrix model 
 including
the  $\chi_{c0} $  resonance with mass $(3414.7\pm 0.3)$ MeV and width $(10.5\pm 0.8)$ MeV~\cite{PDG2019}, suggested to be a tetraquark~\cite{Nielsen:2018ytt}, in the s-wave scattering coupled channels $D\bar{D}-\p\p$.
Furthermore,  focusing on  a mechanism that can generate CP violation in high mass regions, the hadronic charm loop with rescattering is added to a nonresonant amplitude carrying the weak phase, as will be explained and fully explored in what follows.

{\it Decay amplitude model.} A CPV process has to be described by a 
decay amplitude that must have two interfering contributions carrying different strong and weak phases.
 The standard mechanism at quark level to produce CP asymmetry 
 is through the interference of  tree and penguin amplitudes
as proposed in BSS model \cite{BSS}.
In the case of $\bppp$, we assume that the weak phase $\gamma$ come from  the tree level diagram 
with quark transition $b\to u$. For simplicity we  neglect the suppressed penguin contribution $b\to d$ to the direct $\bppp$ decay process.
The hadronic decay channel having as source  tree or loop diagrams at the partonic level can also  contribute with 
a strong phase from the final state interaction or low energy resonances. 
Besides, the  $B$ decay in two charmed mesons have a hadronic penguin like topology, that together with the subsequent rescattering  $D\bar{D}-\p\p$ is assumed to contribute with a strong phase.

\begin{figure}[ht]
\begin{center}
\includegraphics[width=.4\columnwidth,angle=0]{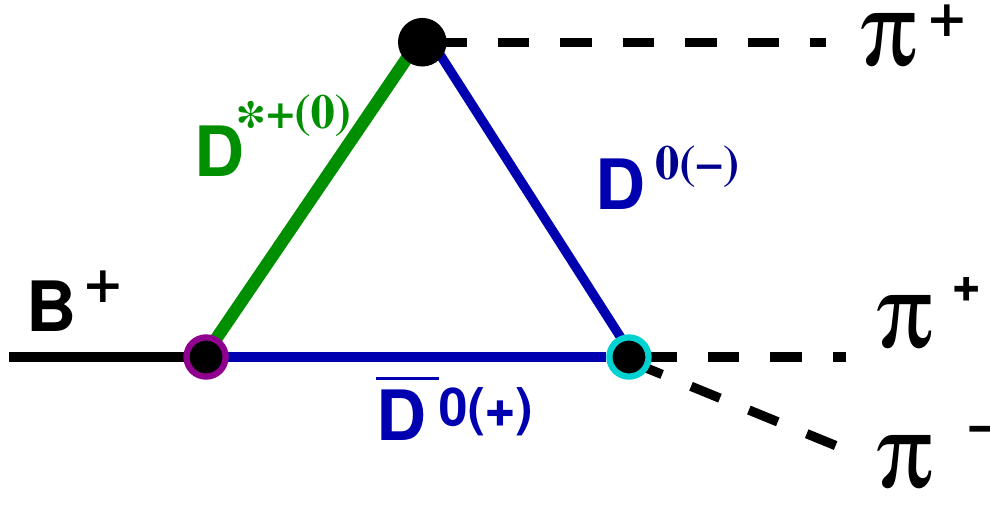}
\caption{ Two different possibilities for the charm loop contribution to $\bppp$ decay. }
\label{fig:tri}
\end{center}
\end{figure}

Inspired by the isobar model description of three-body decays, the amplitude of $\bppp$ decay can be parametrised  by two independent  contributions as:  
\bea
A_{\bppp} (s_{12},s_{23}) = A^{\pm}_{tree}(s_{12},s_{23}) + A_{D\bar{D}}(s_{12},s_{23})\, ,
\label{ampliT}
\eea
where we assume that $A_{D\bar D}$ amplitude is dominated by a charm hadronic loop, Fig.~\ref{fig:tri}, and  $A^\pm_{tree}$ which is the dominant topology, has weak ($\pm \gamma$) and strong phases. Furthermore, the $\chi_{c0}$ will be introduced as a resonant state below threshold within the $D\bar D$ scattering amplitude. 
We will exploit the model in  the high mass region of the $\bppp$ phase space to find out the manifestation in the CP violation distribution
of  the $D\bar D\to \pi\pi$ rescattering,  with $\chi_{c0} $  being a resonant state below the $D\bar D$ threshold. 

A remark on the implication of  CPT invariance  to  CP asymmetry  for the $\bppp$ decay in the present model is appropriate. 
In the framework  developed by Wolfenstein~\cite{Wolfenstein} (see also~\cite{BigiBook}) where the hadronic
final-state interactions and the CPT constraint were considered together, the CP asymmetry seen in channels that can be coupled by strong QCD dynamics
 are related.  The consequence of this framework is that the sum of the partial widths for 
those channels should be identical to the sum in the charge conjugated channels.
Such result is more restrictive than the general CPT condition that gives equal lifetime for a particle and its anti-particle. 
The Wolfenstein formalism was further elaborated in~\cite{Bediaga:2013ela}, where It was considered  the hadronic transition matrix of different channels coupled by FSI 
in the expansion of the CP violating $B$ decay amplitude. 
Restricted to two channels the leading order formalism was applied to study the CP asymmetries seen in the  $\bkkk$ and $\bkpp$ 
in the mass region where the $K^+K^-$ and $\pi^+\pi^-$ channels  are strongly coupled. It explained the remarkable opposite 
 signs and the shape of the projected CP asymmetry.
This mechanism was confirmed by the LHCb collaboration amplitude analyses for $\bkkp$ \cite{BkkpPRL} which found  $65\%$ of asymmetry due to  $KK\to\pi\pi$ with a different sign of the one observed in $B^+ \to \pi^+ \pi^+ \pi^-$ decays~\cite{BpppPRL, BpppPRD}, although with less intensity.

We observe that the leading order formalism also applies to the present model of the three-body $B$ decay where the $B^\pm\to D\bar D \pi^\pm$ and $\bppp$ 
channels are coupled by  the strong force and  the associated  $D\bar D$ and $\pi\pi$  S-matrix provides the FSI contribution to the decay amplitude. The  CP asymmetry of the 
 $B^\pm\to D\bar D \pi^\pm$ has to receive a corresponding contribution with opposite sign respecting CPT invariance if only 
 this channel coupling  is present. However, the $D\bar D$ channel can also coupled to $KK$ as we already 
 discussed in~\cite{Pat2018},  suggesting that the CP asymmetry in $B^\pm\to D\bar D \pi^\pm$ would call for  contributions  from final state interaction
 involving more hadronic channels, a discussion that is much beyond the scope of  the present work.

{\it Hadronic charm loop.}
The charm  rescattering contribution to the $\bppp$ decay can be described by a triangle loop of D mesons as the source for  $D\bar D$, which makes a transition to  $\pi^+\pi^-$, 
 for two possible charge states as depicted in the diagram at Fig.~\ref{fig:tri}.  
In this case,  because both possibilities are similar we consider only the neutral one, $B^+\to D^0\bar{D^0}\p^+$,  which is similar 
to our previous study of the $\bkkk$ decay ~\cite{Pat2017}.

The technique to compute the triangle loop in Fig.~\ref{fig:tri} was already developed in previous works within the context of hadronic 
three-body decays~\cite{Pat2017,Pat2018,PatWV,PatIg}. For the sake of clarity, we repeat some of the  steps required to formulate and compute the loop integral.

We assume factorisation to built the $B\to D\bar{D}\pi$ vertex in the loop diagram which is written as the product:
\begin{equation}\label{Gamma}
\Gamma_{B\to D\bar{D}\pi}= \frac{G_F}{\sqrt{2}}V_{cb}V^*_{cd}\,\la \, \bar{D^0} \, | \, V^\m \, | \, B^+ \, \ra \, g_{\mu\nu}\, \la \, D^0\pi^+ \, | \, V^\nu \, | \, 0 \, \ra ,
\end{equation} 
 where $G_F$  is the Fermi constant and $V_{cb}$ and $V^*_{cd}$ the matrix elements (m.e.) of the CKM matrix.  
The currents m.e.'s are described by  form factors  with the single pole approximation, and for  convenience
we introduce  the notation
$V^\m_{BD}\equiv \la \, \bar{D^0} \, | \, V^\m \, | \, B^+ \, \ra $ 
and $V^\m_{D\pi}\equiv \la \, D^0\pi^+ \, | \, V^\m \, | \, 0 \, \ra $. The former
represents the vector current 
  $B^+\to \bar D^0$ transition and the latter takes into account the amplitude for 
the pair  $D^0\,\p$  produced from the W boson excitation from the vacuum. From crossing
one finds that $V^\m_{D\pi}$ represents the vector current m.e. of the $D\to\pi$ transition obtained from the vector meson dominance (VMD), 
with a single $D^*$ contribution.
 
The vector current m.e. for the transition $B^+\to \bar D^0$  is written as:
\bea
V^\mu_{BD}= \lb p_B^\m + p_{\bar D_0}^\m - \frac{M_B^2 \sm M_{\bar D_0}^2}{\ell^2}\, \ell^\m \rb \, 
F_+(\ell^2)
+ \frac{M_B^2 \sm M_{\bar D_0}^2}{\ell^2}\; \ell^\m  \, F_0(\ell^2) \;,
\label{FFdef}
\eea
 where 
 $\ell = P_B - p_{\bar D_0}$ is the momentum  transfer and the   vector and scalar form factors are, for simplicity, given by:  
 \bea
F_{+(0)}(\ell^2)= - m^2_{B^{*}}\frac{F^{BD}(0)}{\Delta_{B^{*}}(\ell^2)} \, ,
\label{F0}\eea
where $\Delta_{B^*}(k^2)=k^2-m^2_{B^*}+i\epsilon$ follows from describing the form factor as suggested by VMD, with the  coupling of the weak current to $B^*$, namely 
the heavy meson vector ground state with  mass  $m_{B^*}$.

 The  amplitude for the $D^0\,\p$  pair  produced from  the vacuum through the resonance $D^{*}$ 
 is parametrized as:  
\bea
 V^\m_{D\pi}  = \lb p_\p^\m - p_{D^0}^\m - \frac{M_\p^2 - M_{D^0}^2}{\ell^2}\, \ell^\m \rb \, 
F_{D^*}^+(\ell^2)\, ,
\label{FFdefDs}
\eea
where the form factor is $F_{D^*}^+(\ell^2)=m_{D^*}^2F^{D^*}(0)\D_{D^*}^{-1}(\ell^2)$ and  $\D_{D^*} (\ell^2)= \ell^2-  D^*_{pole}$.
The product of both currents  in Eq. (\ref{Gamma}) is written as
\begin{equation}
V^\mu_{BD}V_{\m D\pi}=m_{B^*}^2\,m_{D^*}^2 \,N(\ell,p_\pi;P_B)
 \frac{F^{BD}(0)F^{D^*}(0)}{\Delta_{B^*} (\ell^2)\Delta_{D^*}(\ell^2)}\,,
\end{equation}
where $p_\pi$ is the bachelor pion momentum  and the contraction of the Lorentz structure is given by the invariant:
\begin{equation}\label{N}
N(\ell,p_\pi;P_B)= \D_{D^0}\lp p_{D^0}^2\rp +2\,\D_{\bar{D^0}}\lp p_{\bar D^0}^2\rp - \,2\,s + 3\,M^2_\p + M^2_B - \ell^2 
\end{equation}
where $p_{D^0}=\ell-p_\pi$, $p_{\bar D^0}=P_B-\ell$ and $s=\lp P_B-p_\pi\rp ^2$ is the mass of the pion pair in the transition $D\bar D\to \pi\pi$.  

The full amplitude for the
 hadronic loop  is obtained by integrating the momentum inside the triangle with off-shell propagators, taking into account  the absorptive and dispersive part of the triangle.   And the integral expression is given by:
\begin{equation}
A^B_{D\bar{D}} = i \,C_0 \, T_{D\bar D
\to\pi\pi}(s)\,\int \frac{d^4 \ell}{(2\p)^4} \; 
\frac{N(\ell,p_\pi;P_B)}{\D_{D^0}(p_{D^0}^2) \,\D_{\bar{D^0}}(p_{\bar D^0}^2)  \,\D_{D^*}(\ell^2)\;\D_{B^*}(\ell^2)} \,,
\label{A1.2}
\end{equation}
with 
$$ C_0 = \frac{G_F}{\sqrt{2}}V_{cb}V^*_{cd}\,m_{B^*}^2\,m_{D^*}^2 F^{BD}(0)F^{D^*}(0)\, ,$$
and $T_{\ddpp}(s)$ is  the  $\ddpp$ scattering amplitude, which will be discussed in the sequence.  
Because $T_{\ddpp}(s)$ acts on the s-wave and we assume
minimal unitarity to describe the T-matrix, it is only a function of the invariant mass s and can be factorized out  from the loop integral. 

The loop integral is calculated using the Feynman technique,  which gives: 
\begin{small}
\bea
A^B_{D\bar{D}}=  i C_0 \, T_{D\bar D\to\pi\pi}
\lb R \,\frac{ I_{D_0\bar{D_0}\,B^*} -  I_{D_0\bar{D_0}D^*}}{m^2_{B^*} - D^{*}_{pole}}  -  I_{D_0\bar{D_0}D^*} +   I_{\bar{D_0}D^*B^*}  + 2\, I_{D_0D^*B^*} \rb  ,
\label{Atri}
\eea
\end{small}
where
\begin{equation}
R=  M^2_B + \,M^2_\p  -2 s + M^2_{D_0} + M^2_{\bar{D_0}} - m^2_{B^*}\,,\quad 
 D^*_{pole} = m_{D^*}^2 -i\Gamma_{D^*} \, .
\end{equation}
The functions $I_{xyz}$ are Feynman integrals defined  in \ref{app:1},  which are computed numerically with meson masses and widths from
Ref.~\cite{PDG2019}.

{\it S-matrix and $D\bar D\to\pi\pi$ transition amplitude.} We modify our previous phenomenological amplitude  for $D\bar D \to KK$~\cite{Pat2018} and generalize  it  for $D\bar D \to \pi\pi$. Furthermore,     $\chi_{c0}$ is introduced 
as resonant state below the $D\bar{D}$ threshold. This is an improvement with respect to the previous approach and different from considering only the contribution of 
$\chi_{c0}$ to the $D\bar D \to \pi\pi$ transition as a Breit-Wigner resonance. Generically, a unitary two channel S-matrix can be parametrized as
\begin{equation}\label{sma}
S =  \left( \begin{array}{cc} \eta \, e^{2 i \delta_1} & i\sqrt{1-\eta^2} \, e^{i (\delta_1 + \delta_2)} \\
i\sqrt{1-\eta^2} \, e^{i (\delta_1 + \delta_2)} & \eta \, e^{2 i \delta_2} \end{array} \right)
\end{equation}
where $\delta_1$ and $\delta_2$ are the phase-shifts of  the $\pi\pi$  and $D\bar{D}$  elastic scattering.  $\eta$ is the inelasticity parameter, which accounts for the probability flux between the two channels. Unitarity demands that the off-diagonal S-matrix elements should have a magnitude lower than one, and
its modulus square can be interpreted as the  probability to occur  the transition between the initial and final channels.

We introduce a parametrization for the phase-shifts and inelasticity parameter based on the reasonings presented in \cite{Pat2017,Pat2018,CPV}, brought
to the context of $D\bar D\to \pi\pi$ transition to estimate  $T_{D\bar D\to\pi\pi}(s)$, which is a key ingredient to the hadronic charm loop  to form
 the pions in the final state.
  Of course one should, in principle, resort to the QCD theory to compute the S-matrix, which, is however, much 
 beyond our work. 

A proposal for the dependence  of the transition probability with the two-meson invariant mass, $s$, in light-meson  processes has been discussed in~\cite{CPV},
 in the context of  $P V\to P'X'$ transitions, and here, these qualitative reasonings are brought in the light of  the present case. 
 In a naive description of the $\pi ^+\pi^-$ inelastic collision amplitude, the pions annihilate into a  quark-antiquark pair 
that propagates before  recombining  to produce the heavy-meson pair.  The  intermediate virtual state propagation 
of the quark pair scales roughly  with  the inverse of  Mandelstam invariant $s$. The  breakup of 
the pion into a quark-antiquark pair brings another factor of $s^{-1}$, and similarly for the  formation of the $D$ meson for $s>>m^2_c$ , with $m_c$ the charm 
quark mass. That provides a damping factor of the off-diagonal S-matrix element of $\sim s^{-3}$,  which combined with the
threshold behaviour gives $\sqrt{s-s_{th}}/s^{2.5}$,  keeps the asymptotic form for large $s$. Therefore, we write:
\begin{equation}\label{sinel}
|S_{\pi\pi \to D\bar D}(s)|=\sqrt{1-\eta^2}\sim \mathcal{N}\sqrt{s/s_{thD\bar D}-1}\,/{(s/s_{thD\bar D})^{2.5}}, 
\end{equation}
where the normalization factor $\mathcal{N}$ 
is chosen to keep the modulus of the S-matrix elements  smaller than 1, as required by  the unitarity constraint. 
If we chose $\mathcal{N}=\Lambda^6=(1.24)^6$ in Eq.(\ref{sinel}) then the maximum  value reaches $\sim 0.87$,  
at  $\sqrt{s}\,=\,1.08\, \sqrt{s_{th}}$, which is  close to example of the $s-$wave isospin zero $\pi\pi\to KK$, where the cross section  drops fast and is relevant below $\sqrt{s}\sim$ 1.6 GeV~\cite{cern-munich}. This qualitative formula is also consistent with  one of the possible parametrizations for inelasticity parameter $\eta(s)=\sqrt{1-|S_{\pi\pi\to KK}(s)|^2}$ 
presented in Ref.~\cite{PelPRD05}. 

 The magnitude of the off-diagonal S-matrix element is then written as Eq. (\ref{sinel}), which is valid for $s>s_{thD\bar D}$. 
However, the three-body phase-space for the $B$ decay has two pions below the $D\bar D$ threshold, which
makes necessary the analytic continuation for $s<s_{thD\bar D}$ in the physical sheet of complex momentum, imposing that  $k_2\to i\kappa_2$  with 
$\kappa_2=\tfrac12\sqrt{s_{thD\bar D}-s}$. Furthermore, the amplitude has to be regulated at low values of $s$. One phenomenological 
way  to introduce an infrared cutoff in Eq.  (\ref{sinel}) is:
\begin{eqnarray}\label{absorptionb}
\sqrt{1-\eta^2}=
{\mathcal N} \left( \frac{s}{s_{thD\bar D}}\right)^\alpha\,\sqrt{s/s_{thD\bar D}-1}\,\left({s_{thD\bar D}\over s+s_{QCD}}\right)^{2.5+\alpha} ,
\end{eqnarray}
where $s_{QCD}$ is an infrared scale of QCD estimated to be $s_{QCD}\sim \Lambda_{QCD}^2\sim 0.2$ GeV$^2$.
The factor $s^\alpha$ in the non-physical region, expresses that the coupling between the open channel
and the off-mass-shell $D\bar D$ pair is damped in the non-physical region, but it does not change the large momentum power-law tail
of the amplitude.

Next, we discuss the parametrization of the elastic phase-shift in the $\pi^+\pi^-$ channel that takes the form
dictated by the effective range expansion:
\begin{equation}
e^{2 i \delta_1} = {c + b\,k_1^2 -i k_1\over c + bk_1^2 +i k_1}\, ,
\end{equation}
where  $k_1=\tfrac12\sqrt{s-s_{th\pi\pi}}$ with the respective threshold of
$s_{th\pi\pi}=4M^2_\pi$. The parameters $b=1$~GeV$ ^{-1}$ and $c=0.2$~GeV come from our previous study~\cite{Pat2018}. 

The new aspect of the S-matrix parametrization with respect to our previous work is the introduction of the 
$\chi_{c0}$ as a resonant  state  of the $D\bar{D}$ system below threshold represented  by a pole in diagonal term and to the phase $\delta_2$: 
\bea 
e^{2 i \delta_2}= {(a^{-1})^*-i k_2\over a^{-1}+i k_2}\, ,
\eea
where  $k_2=\tfrac12 \sqrt{s-s_{thD\bar D}}$, and the $D\bar D$ threshold is $s_{thD\bar D}=(M_D+M_{\bar D})^2$.
For $D\bar D$ channel, we choose a complex scattering length dominated parametrization.  We define the scattering length such that the elastic $D\bar D$ amplitude presents 
a pole in the complex plane. The real part below threshold accounts for the  $\chi_{c0}$ mass  and the width moves the pole into the complex plane:
\begin{equation}
a^{-1} - \kappa_{\chi_{c0}}= 0 \quad\text{with}\quad  {\kappa}_{\chi_{c0}} = \frac12\sqrt{ s_{thD\bar D} - M_{\chi_{c0}}^2 + iM_{\chi_{c0}}\Gamma_{\chi_{c0}}  }\, .
\end{equation}

 For $s < s_{thD\bar D}$ the transition amplitude becomes
\begin{equation} 
T_{D\bar D,\pi\pi}(s)= \,
\left( \frac{s}{s_{thD\bar D}}\right)^\alpha\, \frac{2\kappa_2}{\sqrt{s_{thD\bar D}}}\,\left({s_{thD\bar D}\over s+s_{QCD}}\right)^{2.5+\alpha}
F(k_1, i\kappa_2)\, ,
\label{t12c}
\end{equation}
where 
$$ F(k_1,k_2)={\mathcal N}\,\left[\left({ c + bk_1^2-i k_1 \over c + bk_1^2 +i k_1}\right)\,\,
\left({\kappa^*_{\chi_{c0}}-i k_2\over \kappa_{\chi_{c0}}+i k_2}\right)\right]^\frac12\, , $$
with ${\mathcal N}$ a normalization factor. The above formula respects 
the unitarity of the S-matrix model.

For $s\geq s_{thD\bar D}$ the transition amplitude is written as:
\begin{equation} 
T_{D\bar D,\pi\pi}(s)= \, - i\,
\frac{2\,k_2}{\sqrt{s_{thD\bar D}}}\,\,\left({s_{thD\bar D}\over s+s_{QCD}}\right)^{2.5}\, \left({s_{thD\bar D}\over 2 s-s_{thD\bar D}}\right)^{\beta} 
F(k_1,k_2)\, ,
\label{t12d} 
\end{equation}
where $\left({s_{thD\bar D}\over 2 s-s_{thD\bar D}}\right)^{\beta} $ was introduced to modulate the shape of the amplitude bump above the $D\bar D$ threshold as we have 
already used in the study of the $D\bar D\to KK$ \cite{Pat2018}.
Note that the $\chi_{c0}$ pole appears in Eq. (\ref{t12d}) through function $F$, merged with the nonresonant structure of the $\ddpp$ amplitude.
We should observe that our naive power counting can have corrections, and indeed this is the case
as it will be shown in our numerical study. 

\begin{figure}[ht!!!!]
\begin{center}
\hspace*{-2mm}\includegraphics[width=.5\columnwidth,angle=0]{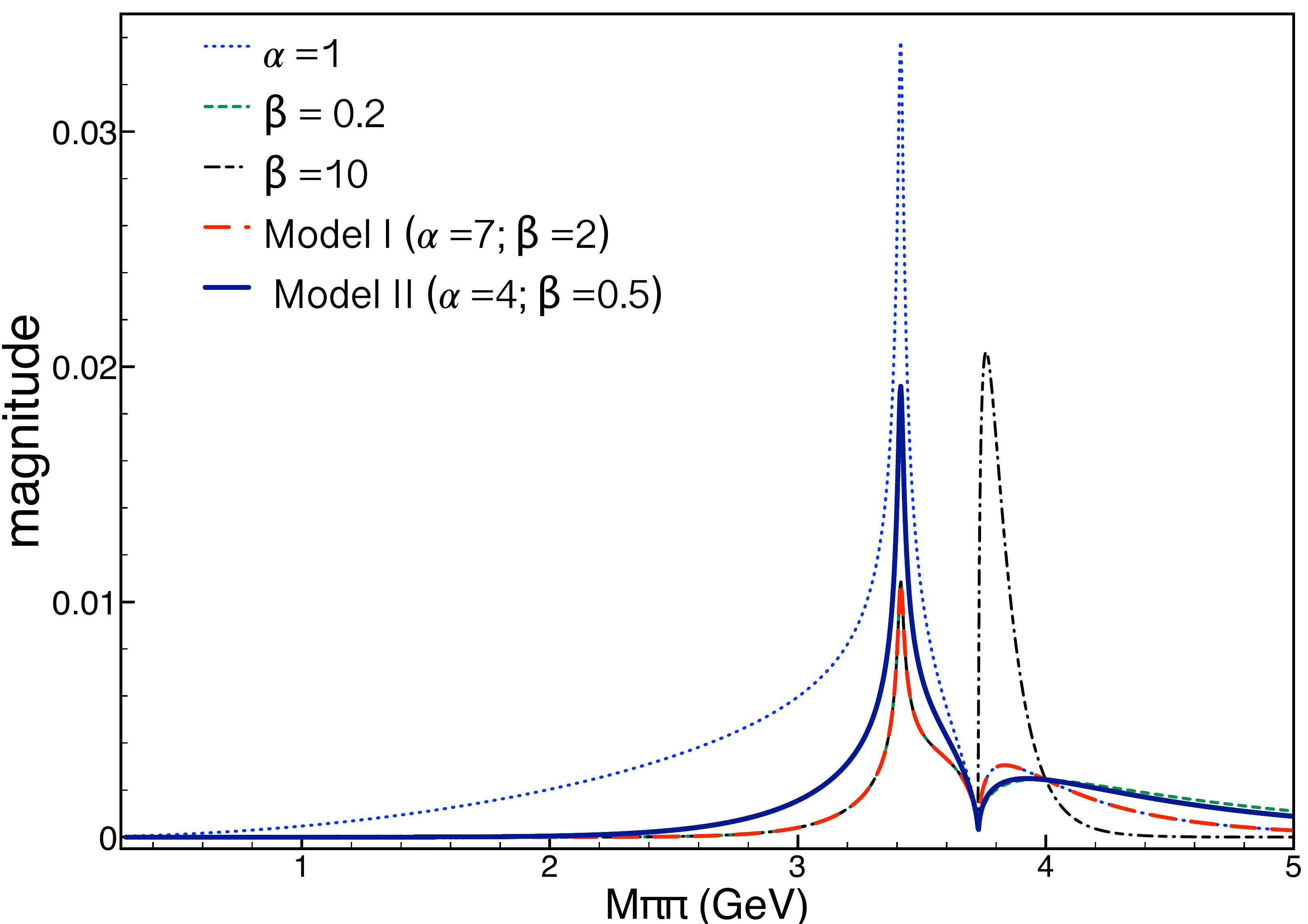}
\includegraphics[width=.5\columnwidth,angle=0]{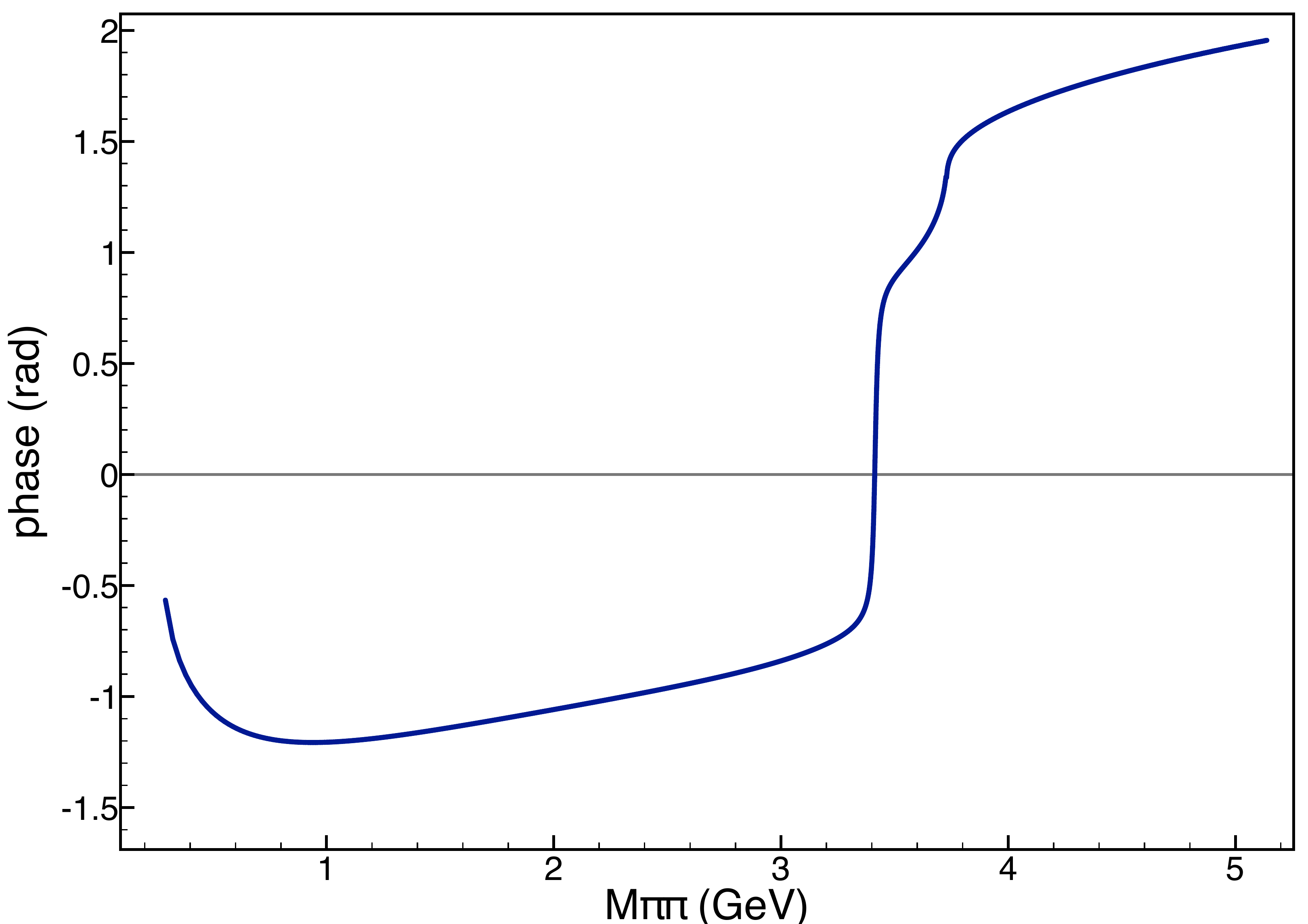}
\caption{ Magnitude and phase of the decay amplitude from the charm hadronic loop with $D\bar D\to \pi\pi$ rescattering,  Eq.~(\ref{A1.2}),
as a function of $m_{\pi\pi}$ (invariant mass of the $\pi\pi$ system). The results for the magnitude are presented for models
I and II  and some variations as indicated within the figure. The phase is not affected by these parameters.}
\label{fig:finalAmpl}
\end{center}
\end{figure}

In our naive modeling we left as free parameters in Eqs.~(\ref{t12c}) and (\ref{t12d}), the exponents $\alpha$ and $\beta$, which can be determined by a fit to the data. 
As a theoretical exercise, we compare the transition amplitude obtained for the same set of parameters found in the study of $\bckkp$ 
(model I: $\alpha =7$ and $\beta =2$) \cite{Pat2018}, and vary $\alpha$ and $\beta$ to find another set (model II:  $\alpha =4$ and $\beta =0.5$), which seems more suitable
to provide a qualitative description of  the experimental data for CPV in the $\bppp$ decay for the high mass region.   
As shown in Fig.~\ref{fig:finalAmpl}, the amplitude from Eqs.~(\ref{t12c}) and (\ref{t12d}) plotted as a function of the $\pi\pi$ invariant mass, 
can have quantitative different signatures depending of the choice of the two exponents, 
but keeping three common features: (i) the $\chi_{c0}$ peak superposed to a wide bump below $D\bar D $  threshold; (ii) the zero at the threshold; (iii) a bump above the
threshold; and (iv) a jump of the strong phase close to $\pi$ when  crossing the $D\bar D$ threshold. The parameters can only move the quantitative values of the 
transition amplitude magnitude, while keeping the qualitative features (i)-(iv).  
The  phase is not affected by the particular choice of  parameters $\alpha$ and $\beta$ once it is connected to the dynamical choice of the amplitude.
We just remind the reader that the $\bppp$ decay amplitude includes the loop integral.

{\it Results for $\bppp$ decay and CPV.} 
The total amplitude model for the $\bppp$ decay, Eq.~(\ref{ampliT}), is the sum of a tree amplitude  $A^{\pm}_{tree}$ and the hadronic charm loop  with $D\bar D\to \pi\pi$ rescattering, $A_{D\bar D}$. Thus the CP asymmetry in $B^\pm$ decays will be the result of the interference between those two.
 In what follows, we are only interested in the dynamics above 3~GeV$^2$  
 where the low mass resonances contributions come mainly from their tails. Therefore,  the amplitude  $A^{\pm} _{tree}$ can be approximated as a flat nonresonant (NR) amplitude with the constant weak phase, $\gamma$:
\bea
A_{tree}^{\pm} = a_0 \,e^{\pm i\gamma}\, ,
\eea
\ni where $a_0$ is complex to accommodate a strong phase.  

The total amplitude was simulated using Laura++ software~\cite{Laura} with hundred thousands  events. 
There are two main variables when two amplitudes interfere: the relative phase between them and the relative magnitude,  in principle those quantities are fixed by a fit to data. 
In our toy model we have to chose $a_0$  and in order to have an insight on the typical results one gets by changing this quantity. We present a systematic study with model II.
 
\begin{figure}[ht!]
\begin{center}
\hspace*{-2mm}\includegraphics[width=.48\columnwidth,angle=0]{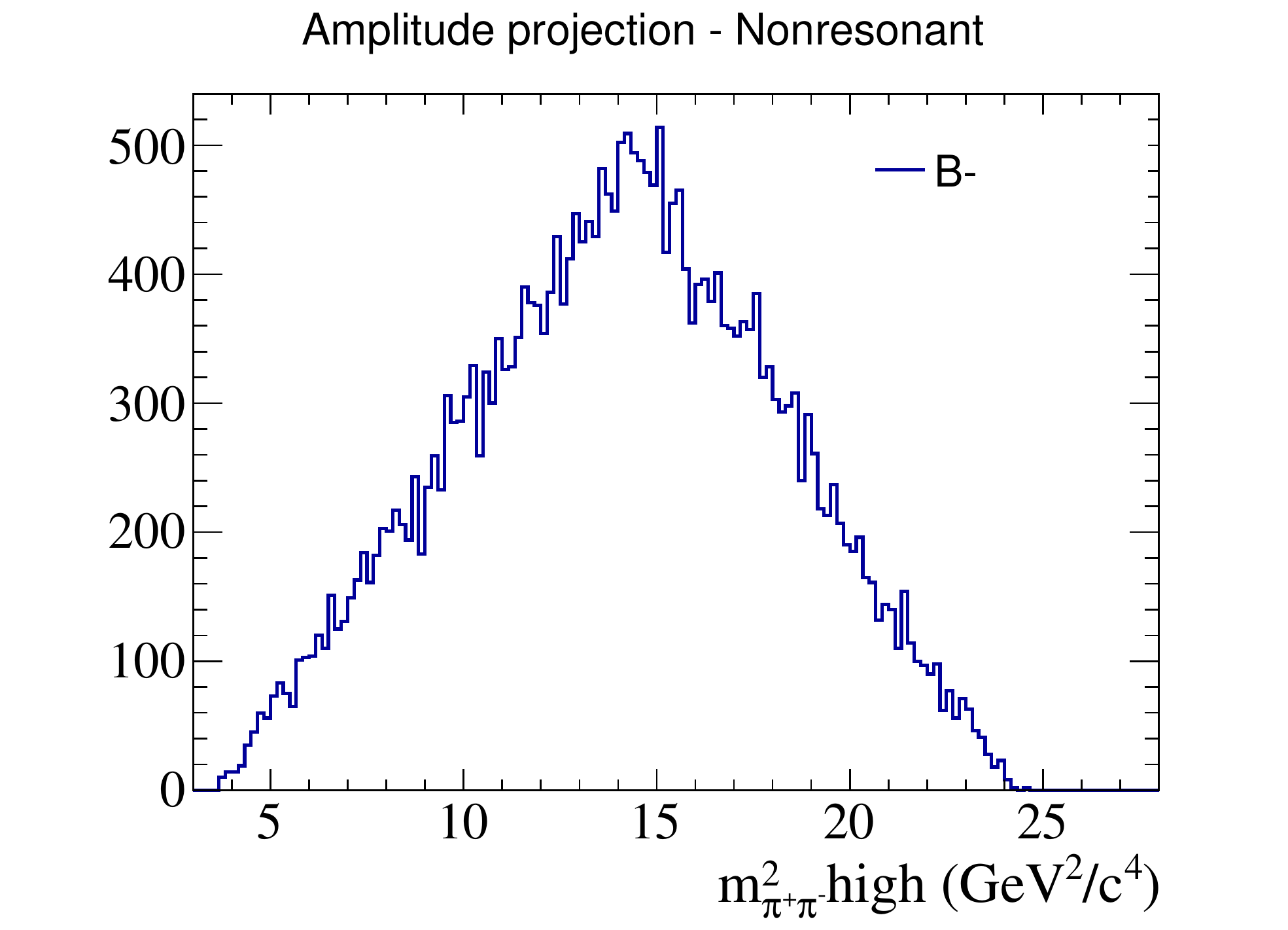}
\hspace*{-2mm}\includegraphics[width=.48\columnwidth,angle=0]{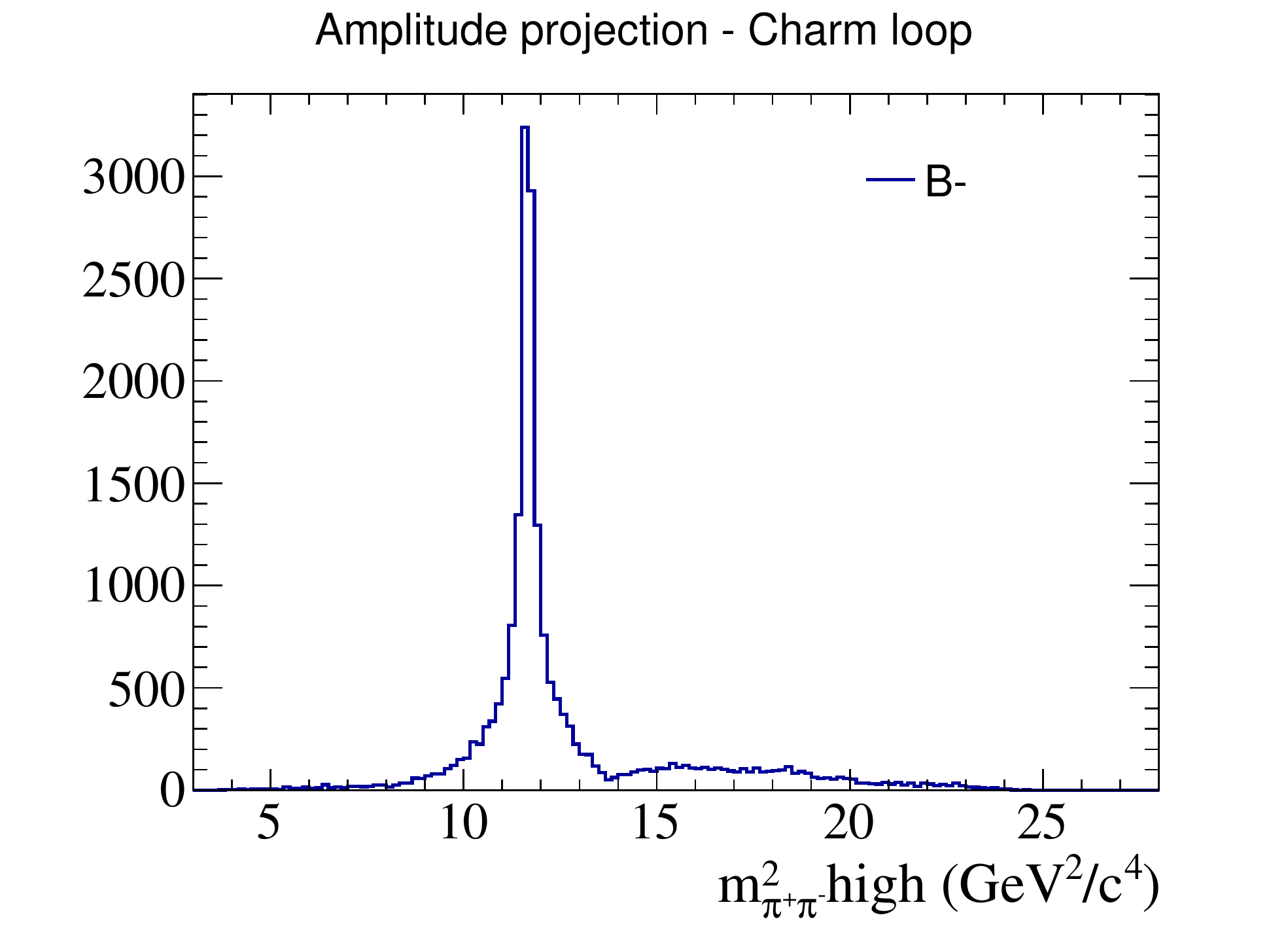}
\caption{LAURA++ Toy Monte-Carlo simulation: (left) only the flat nonresonant tree amplitude; (right) only the charm loop with rescattering amplitude (model II). }
\label{fig:Individual}
\end{center}
\end{figure}

To start our simulations,  it is interesting to check the signature of each amplitude $A^{\pm}_{tree}$ and $A_{D\bar{D}}$ alone in the phase-space projected on the $m_{\p\p}$ high  invariant mass\footnote{defined as the higher one from the two possible pairs of $\p^+\p^-$ invariant masses.}. We integrate in the $m_{\pi\pi}$ low invariant mass starting at 
$m^2_{\pi\pi}$=3~GeV$^2$ to exclude the low energy interaction region. In Fig.~\ref{fig:Individual}, one can see the result  from the flat NR amplitude deformed by the phase-space integral and the hadronic loop with model II.  Each of them alone does not lead to CP violation, as expected.

\begin{figure}[ht!!!!]
\begin{center}
\hspace*{-2mm}\includegraphics[width=.8\columnwidth,angle=0]{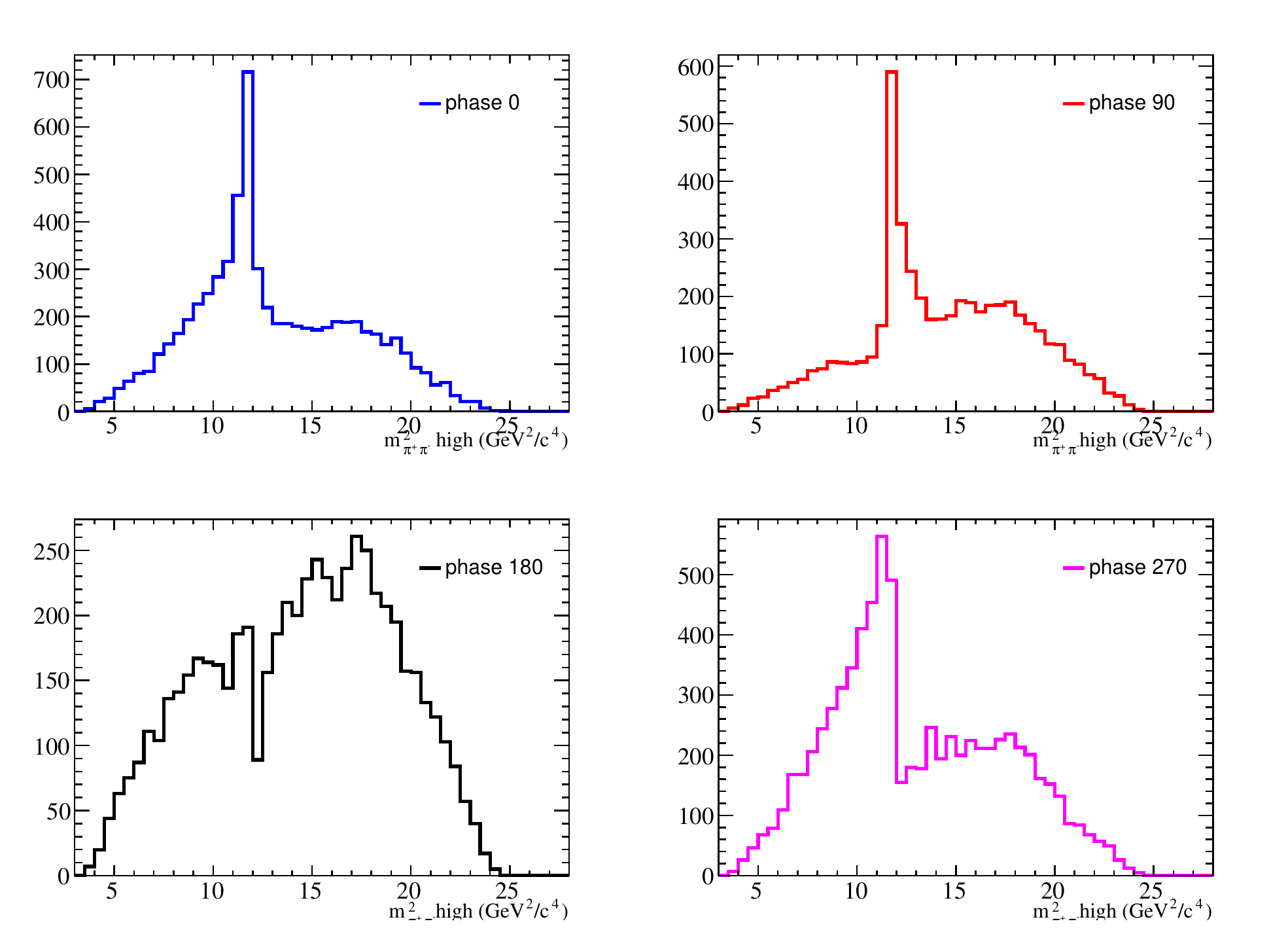}
\caption{Integrated decay rate from the full amplitude (model II) as a function of the $\pi^+\pi^-$ invariant mass. Variation of the relative phase between  $A_{tree}^{\pm}$  and $A_{D\bar D}$
with values taken from $0^o$ up to $270^0$. }
\label{fig:difPhase}
\end{center}
\end{figure}

  In Fig.~\ref{fig:difPhase},  we present the study of  how  the amplitudes interfere with different choices for $a_0$. We set the relative magnitude for the NR to be twice the charm loop and change the relative global phase between them.  As one can see, the different relative phases can result in completely different  patterns, but with a clear
  mark at the resonance position.  In the bottom left frame in Fig.~\ref{fig:difPhase}, the phase difference of $180^o$ eliminates the $\chi_{c0}$ peak and make it appears 
  as a dip. Whereas with $0^o$ phase the peak is enhanced. 

In principle, we have the freedom to chose the relative phase and intensity of the decay amplitudes 
$A^{\pm}_{tree}$ and $A_{D\bar D}$ besides the model parameters, which can be fitted to data. 
However, our goal in this study, is to check if the model is able to reproduce the main characteristics observed in the LHCb data~\cite{LHCb3}: a  
CP asymmetry $({A_{CP}})$
 positive above 3~GeV$^2$ until the region  where the charm channel opens and $A_{CP}$ flips sign. 
We can retrieve such  $A_{CP}$ pattern with model II and a weak phase of $\gamma = 70^o$~\cite{PDG2019} by chosing, guided by the study presented in  Fig.~\ref{fig:difPhase}, 
the relative phase to be $45^o$ with magnitude of the NR  amplitude twice the one for the hadronic charm loop with rescattering.

\begin{figure}[ht!!!!]
\begin{center}
\vspace*{-2mm}\includegraphics[width=.48\columnwidth,angle=0]{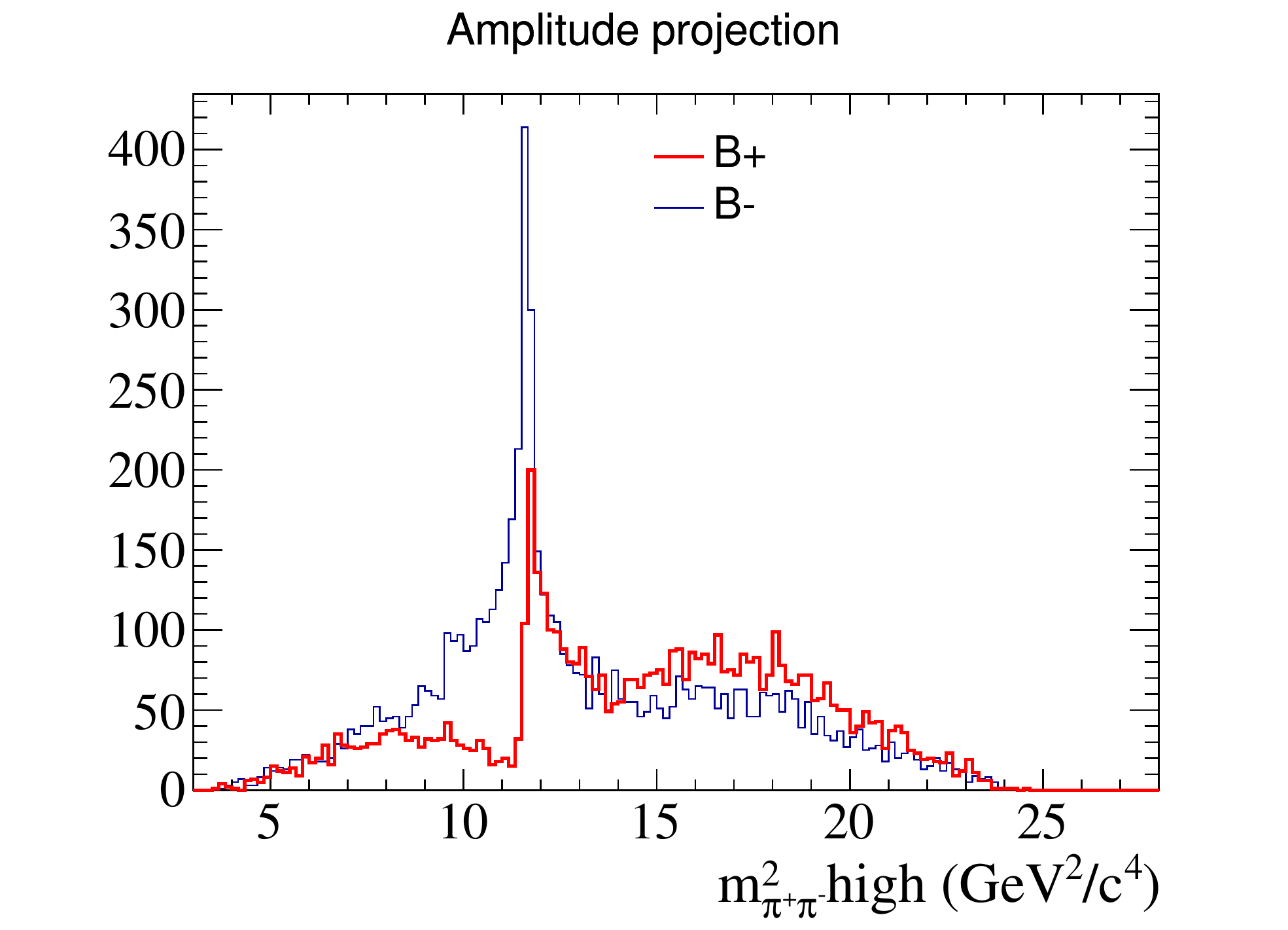}
\includegraphics[width=.48\columnwidth,angle=0]{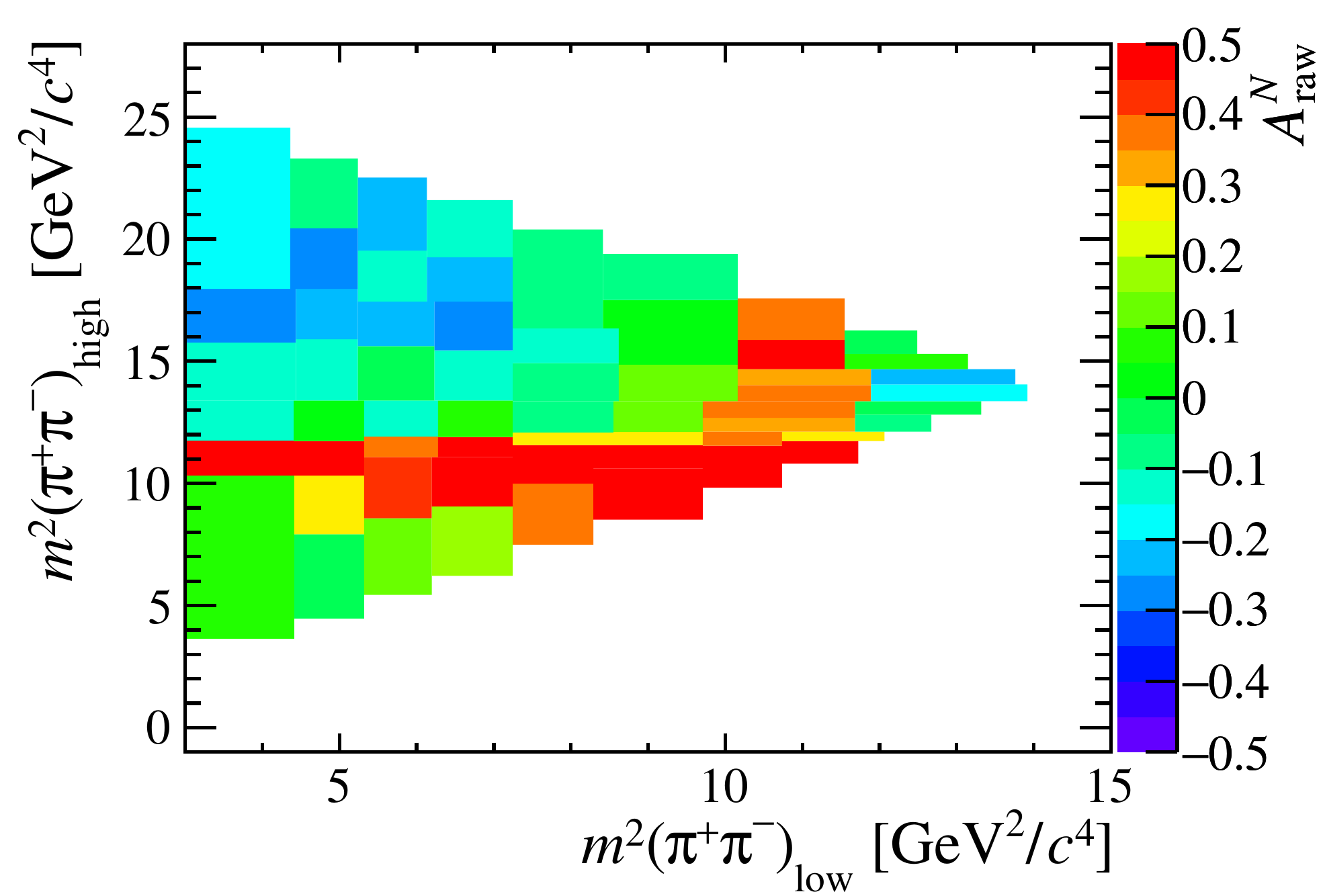} 
\caption{Left: LAURA++ output for the integrated decay rate for model II with the NR amplitude having a strong phase of $45^o$ and weak phase of $70^o$~\cite{PDG2019} with twice the magnitude the charm loop one. Right: Miranda technique~\cite{BedPRD2009} applied to expose the CP violation in different regions of the $\bppp$  phase-space for model II. }
\label{fig: Lauraoutput t12}
\end{center}
\end{figure}

In the left frame of Fig.~\ref{fig: Lauraoutput t12},  we show that we can indeed produce the desired characteristics for $A_{CP}$ described above  for the projection in the three-body phase-space.
We also checked the CP violation signature produced by the interference of the same flat NR amplitude with a simple Breit-Wigner representation 
of the $\chi_{c0}$ in an isobar model with the same relative phase and magnitude as above. We have found that the CP asymmetry is localized in a  much smaller region around 
$\chi_{c0}$ compared what we have observed with the  rescattering model.

In order to study the CPV signature between the  $B^+$ and $B^-$ 
in the high mass region, 
 we use the Miranda technique~\cite{BedPRD2009} and present the CPV distribution in three-body phase-space on the right frame of Fig.~\ref{fig: Lauraoutput t12}. This can be compared to the same CPV Dalitz plot figure produced by LHCb data for $\bppp$ decays~\cite{LHCb3}.   From the projection in Fig.~\ref{fig: Lauraoutput t12}  it is clear the signature of the $\chi_{c0}$ peak coming from the $D\bar D$ resonant state below threshold 
 widened by the $D\bar D\to\pi\pi$ rescattering.
Whereas from the Dalitz plot in  Fig.~\ref{fig: Lauraoutput t12}, one can see the red band for positive $CP$ asymmetry in the  $\chi_{c0}$ region followed by a blue band pointing to a change of sign around the $D\bar D$ threshold. A similar pattern can be identified in the experimental data~\cite{LHCb3}. We recall that there are other contributions  that could spread the CP asymmetry  of the $\bppp$ decay in the high mass region, 
which were not considered here, like
the tails of the low mass resonances, the excited states of the $D$ systems, still coupled to $\pi\pi$ channels,
and/or three-body rescattering in the $D\bar D \pi$ channel.

{\it Summary.}  We developed a model for the $\bppp$ decay amplitude, which has contribution from a tree $b\to u$ nonresonant amplitude and  a hadronic charm loop  with a s-wave $D\bar D\to \pi\pi$ rescattering, 
where  $\chi_{c0}$  is introduced as a resonant state of the $D\bar D$ system below threshold with the narrow experimental width. 
The  $\chi_{c0}$  pole of the elastic $D\bar D$ scattering amplitude modifies the $D\bar D\to \pi\pi$ transition amplitude due 
to the assumed S-matrix unitarity of the two-channel model. 
With this simple  model for $\bppp$ decay amplitude we were able to mimic qualitatively the  CP asymmetry distribution reported by LHCb Run I data in the high mass region~\cite{LHCb3}, giving a possible interpretation of the mechanism behind these challenging experimental results. Therefore, we strongly encourage the experimentalists to incorporate the present model in their amplitude analyses for the next data generation in order to improve our understanding 
of the nature of CP violation in charmless three-body $B$ decays in the high mass region.

\section*{ Acknowledgements} 
We would like to thanks Jussara Miranda and Keri Vos for their fruitful discussions. 
PCM work was supported by  Marie Currie (MSCA) grant no. 799974. 
This work was partly supported by the Funda\c c\~ao de Amparo \`a Pesquisa do Estado de
 S\~ao Paulo [FAPESP grant no. 17/05660-0],
Conselho Nacional de Desenvolvimento Cient\'ifico e Tecnol\'ogico [CNPq grants no.
 308025/2015-6, 308486/2015-3 ] and
Coordena\c c\~ao de Aperfei\c coamento de Pessoal de N\'ivel Superior (CAPES) of Brazil. 
This work is part of the project INCT-FNA Proc. No. 464898/2014-5.

\appendix
\section{Charm loop integrals} \label{app:1}
\begin{figure}[ht!!!!]
\begin{center}
\vspace*{-2mm}\includegraphics[width=.35\columnwidth,angle=0]{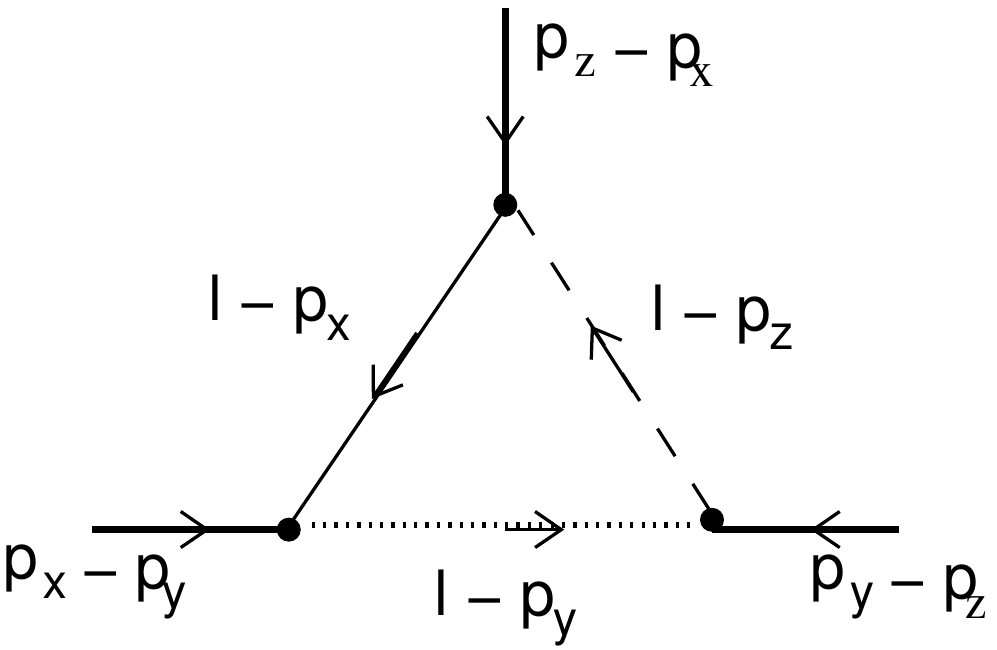}
\caption{ A triangle loop integral. }
\label{tri}
\end{center}
\end{figure}

A general  triangle loop integral is written as the following form:
\bea
I_{x y z} =
\int \! \frac{d^4 l}{(2\p)^4} \;
\frac{1}{[(p_x \sm l)^2 \sm m_x^2 \sp i\,\e]\;} 
\frac{1}{[(p_y \sm l)^2 \sm m_y^2 \sp i\,\e]}\;\frac{1}{
[(p_z \sp l)^2 - m_z^2\,]} \;, 
\label{pu.1}
\eea
where the momenta $p_x, \, p_y$ and $p_z$ are shown in  Fig.~\ref{tri} for the representation of the loop diagram.

The loop integral can be done using the standard Feynman technique:
\bea
I_{xyz} =- \frac{i}{(4\p)^2}\,\int_0^1 da\; a \, \int_0^1 db\; \frac{1}{D_{x y z}}\;,
\label{triangulo.1}\eea  
where the denominator is given by:
\bea
D_{xyz} &=& \bar a\,m^2_x + a\bar b\,m^2_y + ab\, m^2_z -a \bar a\bar b(p_x -p_y)^2 \nn \\ && 
- a \bar a b\,(p_x -p_z)^2- a^2 b \bar b\,(p_y -p_z)^2 - i\e\, ,
\label{Dxyz}
\eea
where $\bar a=1-a$ and $\bar b=1-b$.

For the specific case of $\bppp$ the four independent functions in Eq.~(\ref{Atri}),
 $I_{D_0\bar{D_0}\,B^*}$,   $I_{D_0\bar{D_0}D^*}$,    $I_{\bar{D_0}D^*B^*}$  and $I_{D_0D^*B^*}$,
 are obtained from the numerical integration of Eq.~(\ref{triangulo.1}), with the denominators written explicitly as:

\bea
 D_{D_0\bar{D_0}\,B^*} &=& 
M^2_B\,(a\,b)^2 + a\,b \lp m^2_{B^*}  - M_{D_0}^2 + \bar a (s - M^ 2_\pi) - a\,M_B^2 \rp \nn \\ &&+   \bar a\,M_{\bar{D_0}}^2 + a\,M_{D_0}^2 -\bar a a\,s - i\e\, ,
\\[2mm]
D_{D_0\bar{D_0}D^*} &=&M^2_B(a\,b)^2 + a\,b\,(D^*_{pole}  - M_{D_0}^2 + \bar a(s - M^ 2_\pi) - a\,M_B^2 )  \nn \\ &&+ \bar a\,M_{\bar{D_0}}^2 + a\,M_{D_0}^2 -\bar a a\,s- i\e\, ,
 \\[2mm]
 D_{\bar{D_0}D^*B^*} &=& a\,b ( D^*_{pole} - m^2_{B^*} )  + a\,m^2_{B^*}   + \bar aM_{\bar{D_0}}^2 - a\bar a\,M_B^2- i\e\,  ,
 \\[2mm]
 D_{D_0D^*B^*} &=& a\,b ( D^*_{pole} - m^2_{B^*} )  + a\,m^2_{B^*}   + \bar a M_{D_0}^2 - a\bar a\,M_{\pi}^2- i\e\, ,
\eea
and for the numerical integration we use a finite value of $\epsilon=0.01$ GeV.

\end{document}